\newcommand{\Msol}{\mbox{$M_{\sun}$}}

\def\MeV{\mbox{Me\hspace{-0.1em}V}}

\def\sun{\hbox{$\odot$}}

\documentclass{elsart}
\usepackage{epsfig}
\begin{document}

%%%%%%%%%%%%%%%%%%%%%%%%%%%%%%%%%%%%%%%%%%%%%%%%%%%%%%%%%%%%%%%%%%%%%%%%%%%%%%
% Titlepage
%%%%%%%%%%%%%%%%%%%%%%%%%%%%%%%%%%%%%%%%%%%%%%%%%%%%%%%%%%%%%%%%%%%%%%%%%%%%%%
\runauthor{Kn\"odlseder}
\begin{frontmatter}
\title{Constraints on stellar yields and SNe from gamma-ray line observations}
\author[jurgena,jurgenb]{J\"urgen Kn\"odlseder}
\address[jurgena]{INTEGRAL Science Data Centre, Chemin 
d'Ecogia 16, 1290 Versoix, Switzerland}
\address[jurgenb]{Centre d'Etude Spatiale des Rayonnements, 31028 
Toulouse Cedex 4, France}

%%%%%%%%%%%%%%%%%%%%%%%%%%%%%%%%%%%%%%%%%%%%%%%%%%%%%%%%%%%%%%%%%%%%%%%%%%%%%%
% Abstract
%%%%%%%%%%%%%%%%%%%%%%%%%%%%%%%%%%%%%%%%%%%%%%%%%%%%%%%%%%%%%%%%%%%%%%%%%%%%%%
\begin{abstract}
Gamma-ray line observations provide a versatile tool for studies of
nucleosynthesis processes and supernova physics.
In particular, the observation of radioactive species in the interstellar 
medium probes recent nucleosynthesis activity on various time-scales 
for different kinds of sources.
Considerable progress in gamma-ray instrumentation during the last decades 
has led to the discovery of several cosmic gamma-ray lines.
In this review, recent observational results are presented and their 
astrophysical implications are discussed.
Prospects of gamma-ray line astronomy will be explored in view of the 
future INTEGRAL mission.
\end{abstract}
\end{frontmatter}

%%%%%%%%%%%%%%%%%%%%%%%%%%%%%%%%%%%%%%%%%%%%%%%%%%%%%%%%%%%%%%%%%%%%%%%%%%%%%%
% Introduction
%%%%%%%%%%%%%%%%%%%%%%%%%%%%%%%%%%%%%%%%%%%%%%%%%%%%%%%%%%%%%%%%%%%%%%%%%%%%%%
\section{Introduction}

During the last decade the field of gamma-ray line astronomy has made 
important progress.
On the one hand, the explosion of the nearby supernova SN 1987A in the 
Large Magellanic Cloud provided us with a bright source of nuclear
gamma-ray lines due to the decay of freshly produced radioactive 
isotopes.
On the other hand, a new generation of space borne telescopes, such as 
COMPTEL and OSSE on the {\em Compton Gamma-Ray Observatory},
and TGRS on {\em WIND}, provided for the first time sufficient sensitivity,
angular and spectral resolution for a comprehensive study of 
galactic gamma-ray lines.
Still, gamma-ray line astronomy is plagued by a dominating 
instrumental background, induced by cosmic-ray bombardment in the 
hostile space environment, reducing line detections to 
significancies below typically $5-10~\sigma$.
Consequently, uncertainties on observed line fluxes or profiles are 
still quite important.
Nevertheless, the available observational material is continuously 
growing, starting to provide interesting constraints on nucleosynthesis 
processes.

Most gamma-ray lines are produced in desexcitation transitions
between nuclear levels.
These gamma-ray lines may be subdivided into 2 categories that are defined 
by the channel that led to the nuclear excitation: 
(a) radioactive decay into an excited state of the daughter isotope, 
or (b) nuclear interactions and reactions.
Radioactive decays are eventually accompanied by positron emission,
resulting in a 511 keV gamma-ray line from e$^{+}$e$^{-}$ 
annihilation.
Other mechanisms leading to positron production involve compact 
objects, where high densities or strong magnetic fields favour 
e$^{+}$e$^{-}$ pair production.

%%%%%%%%%%%%%%%%%%%%%%%%%%%%%%%%%%%%%%%%%%%%%%%%%%%%%%%%%%%%%%%%%%%%%%%%%%%%%%
% Radioactivity lines
%%%%%%%%%%%%%%%%%%%%%%%%%%%%%%%%%%%%%%%%%%%%%%%%%%%%%%%%%%%%%%%%%%%%%%%%%%%%%%
\section{Radioactivity lines}

Gamma-ray lines from radioactivities demand several conditions to be 
observable.
First, a hot and dense medium with sufficiently low entropy is required to 
allow for the synthesis of fresh radioisotopes.
Such a medium can be found in stellar interiors, at the base of the 
accreted envelope of white dwarfs in close binary systems, or even in 
accretion disks around compact objects.
The nuclear reaction networks in operation are characteristic for the 
composition, density, and temperature at the burning site, hence 
the observation of isotopic abundance patterns provide direct insight 
into the nucleosynthesis conditions.
Second, the fresh radioisotopes have to be removed quickly from the 
formation site to prevent destruction by nuclear reactions or natural 
decay.
This generally implies convection followed by mass ejection, either 
in form of stellar winds or explosions, and requires lifetimes of at 
least several days, better several months.
Additionally, nucleosynthesis sites are generally optical thick to 
gamma-rays, hence escape of the radioisotopes to optically thin 
regions is mandatory for gamma-ray line observations.
Consequently, radioisotopes can probe stellar convection and ejection 
processes, providing important information about the involved stellar 
physics.
Third, the lifetime has to be short enough and the abundance of the 
isotope has to be high enough to assure a sufficient radioactive 
decay activity that is in reach of modern gamma-ray telescopes.

%%% Table: Gamma-ray lines from radioactivities %%%%%%%%%%%%%%%%%%%%%%%%%%%%%%
\begin{table*}
\caption{Gamma-ray lines from radioactivities that may be accessible 
         to gamma-ray astronomy (ordered by ascending lifetime).
         Theoretical nucleosynthesis yield estimates are quoted for 
         different source types.
         Positron emitters are marked by $\dagger$.}
\label{tab:radioactivities} 
\begin{center}
\begin{tabular}{c c c c c c c c}
\hline 
Isotope & Lifetime $\tau$ & Lines (keV) & \multicolumn{5}{c}{Typical 
yields (\Msol)} \\
\cline{4-8}
& & & WR & SN Ia & SN Ib/c & SN II & Nova \\
\hline
\nuc{57}{Ni} & 2.14 d     & 1378         & 
& 0.02 & $5\,10^{-3}$ & $5\,10^{-3}$ & \\	
\nuc{56}{Ni} & 8.5 d      &	158, 812 &
& 0.5 & 0.1 & 0.1 & \\	
\nuc{59}{Fe} & 64.2 d     &	1099, 1292    &
& & $5\,10^{-5}$ & $5\,10^{-5}$ & \\	
\nuc{7}{Be}	 & 77 d       & 478          &
& & $10^{-7}$ & $5\,10^{-7}$ & $5\,10^{-11}$ \\	
\nuc{56}{Co}$^{\dagger}$ & 112 d      &	847, 1238 &	
& 0.5 & 0.1 & 0.1 & \\	
\nuc{57}{Co} & 392 d      &	122          &
& 0.02 & $5\,10^{-3}$ & $5\,10^{-3}$ & \\	
\nuc{22}{Na}$^{\dagger}$ & 3.76 yr    &	1275         &
& $10^{-8}$ & $10^{-6}$ & $10^{-6}$ & $5\,10^{-9}$ \\	
\nuc{60}{Co} & 7.61 yr    & 1173, 1333    &
& & $10^{-5}$ & $10^{-5}$ & \\				
\nuc{44}{Ti}$^{\dagger}$ & 87 yr      & 1157   &
& $10^{-5}$ & $5\,10^{-5}$ & $5\,10^{-5}$ & \\	
\nuc{26}{Al}$^{\dagger}$ & $10^6$ yr  &	1809         &
$10^{-4}$ & & $5\,10^{-5}$ & $5\,10^{-5}$ & $10^{-8}$ \\	
\nuc{60}{Fe} & $2.2\,10^6$ yr &	1173, 1333 &	
& $5\,10^{-3}$ & $5\,10^{-5}$ & $5\,10^{-5}$ & \\	
\hline 
\end{tabular}
\end{center}
\vspace*{.6cm}
\end{table*}
%%%%%%%%%%%%%%%%%%%%%%%%%%%%%%%%%%%%%%%%%%%%%%%%%%%%%%%%%%%%%%%%%%%%%%%%%%%%%%

These constraints result in a list of candidate isotopes that may 
actually be accessible to gamma-ray line astronomy (cf. Table
\ref{tab:radioactivities}).
The observation of radioactivity lines implies various 
time-scales.
For lifetimes that are short compared to the event frequency
(\nuc{57}{Ni} - \nuc{57}{Co}), individual {\em transient} gamma-ray line 
sources are expected, mainly in form of supernovae or novae.
For lifetimes of the order of the event frequency
(\nuc{22}{Na} - \nuc{44}{Ti}) several rather steady gamma-ray line 
sources are expected in from of supernova remnants or recent nova 
events.
For lifetimes that are long compared to the event frequency 
(\nuc{26}{Al} - \nuc{60}{Fe}) a superposition of numerous individual 
gamma-ray line sources will lead to a diffuse glow of gamma-ray line 
emission.
Additionally, the radioisotopes may travel considerable distances away 
from the production sites before they decay ($\sim10-100$ pc), leading 
to intrinsically extended sources.
Hence, diffuse galaxywide emission is expected for \nuc{26}{Al} and
\nuc{60}{Fe} (see \cite{pd96,diehl98} for recent reviews).

%%%%%%%%%%%%%%%%%%%%%%%%%%%%%%%%%%%%%%%%%%%%%%%%%%%%%%%%%%%%%%%%%%%%%%%%%%%%%%
\subsection{SN 1987A - a nucleosynthesis laboratory}

The explosion of SN 1987A in the Large Magellanic Cloud was a great 
opportunity for gamma-ray line astronomy.
For the first time, a supernova explosion occurred close enough to be
in reach of available gamma-ray telescopes.
During core collapse, substantial amounts of \nuc{56}{Ni} and 
\nuc{57}{Ni} are produced which subsequently decay under gamma-ray lines 
emission to \nuc{56,57}{Co} and finally to \nuc{56,57}{Fe} 
(cf.~Table \ref{tab:radioactivities}).
The production of these isotopes in supernova explosions has been 
indirectly inferred from lightcurve characteristics, reflecting the 
respective decay times.
The direct observation of the gamma-ray lines from \nuc{56}{Co} 
\cite{matz88} and \nuc{57}{Co} \cite{kurfess92} in SN 1987A was a 
brilliant confirmation of this interpretation.
The observed relative intensities of the gamma-ray lines from \nuc{56}{Co} and 
\nuc{57}{Co} indicated a \nuc{57}{Ni}/\nuc{56}{Ni} ratio
between $1.5 - 2$ times the solar ratio of \nuc{57}{Fe}/\nuc{56}{Fe}, 
consistent with core collapse supernova models \cite{woosley91}.

Surprisingly, the \nuc{56}{Co} lines were detected already 6 months after 
explosion, at an epoch where standard onion-shell supernova expansion models 
still predicted a substantial gamma-ray opacity for the envelope.
The gamma-ray line lightcurves presented clear evidence that \nuc{56}{Co} 
was found over a large range of optical depths, with a small 
fraction at very low depth \cite{leising90}.
Probably some fragmentation of the ejecta and acceleration of the 
emitting radioactivity within are required to explain the observations.
The acceleration hypothesis is supported by various gamma-ray line 
profile measurements all indicating line widths of order 1 \% FWHM, 
corresponding to Doppler velocities of 3000 km~s$^{-1}$
\cite{mahoney88,rester89,tueller90}.

Measurements of the bolometric SN 1987A lightcurve indicate that also 
some \nuc{44}{Ti} has been produced during the explosion.
The expected 1.157 MeV gamma-ray line intensities are actually too 
weak for current telescopes, but the decay of \nuc{44}{Ti} is 
sufficiently slow that it will be observable by future, more sensitive 
instruments.
Hence, SN 1987A still remains an interesting nearby laboratory for 
studies of explosive nucleosynthesis processes.

%%%%%%%%%%%%%%%%%%%%%%%%%%%%%%%%%%%%%%%%%%%%%%%%%%%%%%%%%%%%%%%%%%%%%%%%%%%%%%
\subsection{\nuc{44}{Ti} - unveiling recent supernova}

The census of recent galactic supernova events is exclusively based on historic 
records of optical observations and amounts to 6 events during the 
last 1000 years.
Due to galactic absorption and observational bias, this census is by 
far not complete.
Gamma-ray line observations of the \nuc{44}{Ti} isotope have the 
potential to considerably increase the statistics.
\nuc{44}{Ti} is believed to be exclusively produced by supernova events, 
and its production can be inferred from the solar abundance of the 
decay product \nuc{44}{Ca}.
Due to the penetrating power of gamma-rays, \nuc{44}{Ti} lines from recent 
supernova events throughout the Galaxy can reach the Earth, and 
therefore unveil yet unknown young supernova remnants.

The prove of principle was recently achieved by the first observation 
of a 1.157 \MeV\ gamma-ray line from the 320 years old Cas A supernova 
remnant using the COMPTEL telescope \cite{iyudin94}.
Note that Cas A almost escaped visual detection if it were not 
accidentally included in Flamsteed's sky atlas as a star of 6th 
magnitude.
This fact is indeed very puzzling since nucleosynthesis models 
predict large amounts of \nuc{56}{Ni} being co-produced with 
\nuc{44}{Ti}, which would have resulted in a visual peak magnitude of 
-4$^{m}$ for Cas A.
Although a possible extinction of 10 mag could solve this problem, it 
is yet to be confirmed observationally.
Alternative explanations invoke ionisation of \nuc{44}{Ti} or 
an asymmetric explosion.
Ionisation would prevent \nuc{44}{Ti} decaying by orbital electron 
capture and hence falsify the relation between observed gamma-ray 
line flux and present \nuc{44}{Ti} mass, relaxing the constraint on 
the amount of \nuc{56}{Ni} produced during the explosion 
\cite{mochizuki99}.
An asymmetric explosion could increase \nuc{44}{Ti} over \nuc{56}{Ni} 
production in the high entropy alpha-rich freezeout along the polar 
directions \cite{nagataki98}.

Evidence for another galactic \nuc{44}{Ti} source was recently found 
in the Vela region where no young supernova remnant was known before 
\cite{iyudin98}.
Triggered by this discovery, a re-analysis of ROSAT X-ray data indeed 
revealed a spherical structure at the position of the new \nuc{44}{Ti} 
source, now identified as the RX J0852.0-4622 supernova remnant 
\cite{aschenbach98}.
Although the \nuc{44}{Ti} observation is only marginal 
\cite{schoenfelder99}, it is the first time that gamma-ray line 
observations triggered the discovery of a new supernova remnant.
Again, no optically bright display has been recorded for RX 
J0852.0-4622 -- are \nuc{44}{Ti} producing supernovae optically 
faint (or obscured)?

%%%%%%%%%%%%%%%%%%%%%%%%%%%%%%%%%%%%%%%%%%%%%%%%%%%%%%%%%%%%%%%%%%%%%%%%%%%%%%
\subsection{\nuc{26}{Al} - recent galactic star formation history}

Intense galactic gamma-ray line emission at 1.809 \MeV, attributed to 
the radioactive decay of \nuc{26}{Al}, has been reported by numerous 
instruments (see \cite{pd96} for a review).
In principle, \nuc{26}{Al} could be produced in appreciable amounts by a 
variety of sources, such as 
massive mass losing stars (mainly during the Wolf-Rayet phase), 
Asymptotic Giant Branch stars (AGBs),
novae (mainly of ONe subtype),
and core collapse supernovae.
Considerable uncertainties that are involved in the modelling of 
nucleosynthesis processes, mainly due to the poorly known physics of 
stellar convection, do not allow for a theoretical determination of 
the dominant galactic \nuc{26}{Al} sources.

\begin{figure*}
\epsfxsize=9.2cm \epsfclipon
\epsfbox{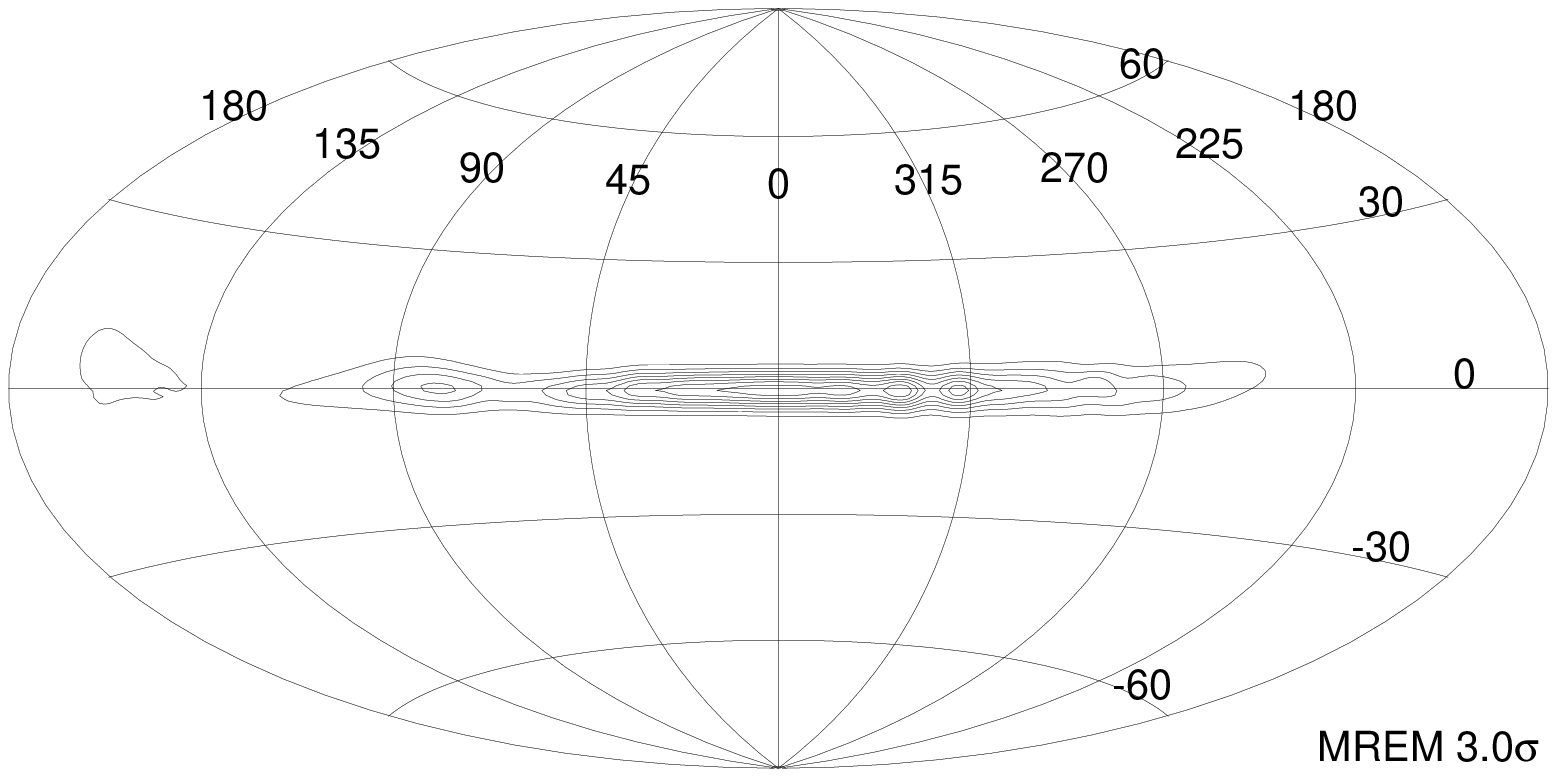}
\hfill
\epsfxsize=4.6cm \epsfclipon
\epsfbox{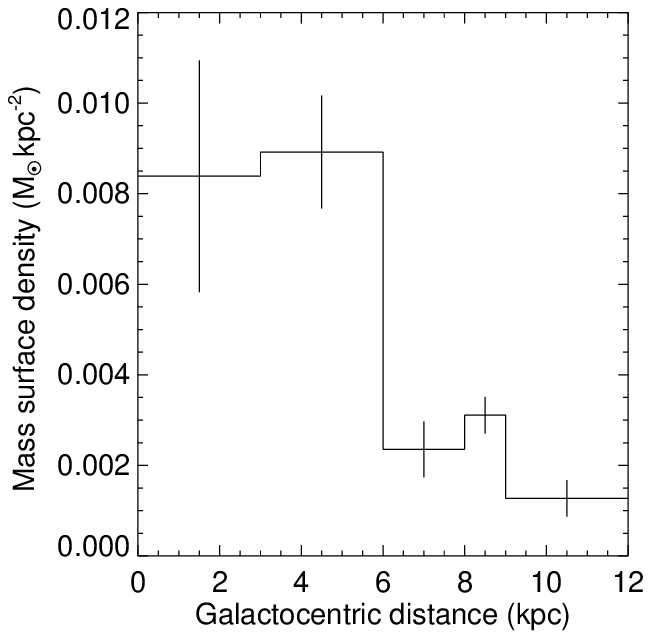}
\caption{{\em Left:} COMPTEL 1.809 \MeV\ gamma-ray line allsky map 
         \cite{knoedl99b}.
         {\em Right:} Radial \nuc{26}{Al} mass density profile 
         \cite{knoedl98}.}
\label{fig:al26}
\end{figure*}

The 1.809 \MeV\ gamma-ray line has now for the first time being imaged 
using the COMPTEL telescope 
\cite{knoedl94,diehl95,oberlack96,knoedl99b}.
The COMPTEL image shows an intense, asymmetric ridge of diffuse galactic 
1.809 \MeV\ emission with a prominent localised emission enhancement 
in the Cygnus region (cf.~Fig.~\ref{fig:al26}).
Additional hints for emission peaks along the galactic plane can be 
understood as fingerprints of the galactic spiral pattern.
Globally, the distribution of 1.809 \MeV\ gamma-ray line emission 
follows very closely the distribution of galactic free-free emission
\cite{knoedl99c}.
Since galactic free-free emission is an excellent tracer of the 
massive star population (M$_{i} > 20$ \Msol), the close correlation 
suggests that \nuc{26}{Al} is mainly produced by this population
\cite{knoedl99a}.
Consequently, due to the short lifetime of massive stars, \nuc{26}{Al} 
becomes an excellent tracer of recent galactic star formation.

The radial \nuc{26}{Al} mass density distribution illustrates that 
the bulk of galactic star formation occurs at distances of less than 6 kpc 
from the galactic centre (cf.~Fig.~\ref{fig:al26}).
Star formation is also present within the central 3 kpc of the Galaxy, 
although at a poorly determined rate.
There are indications for enhanced star formation between $3-6$ kpc, 
coinciding with the molecular ring structure as seen in CO data 
\cite{dame87}.
Enhanced star formation is also seen in the solar neighbourhood 
($8-9$ kpc) which probably corresponds to the local spiral arm 
structure.
However, the radial \nuc{26}{Al} profile is probably not directly 
proportional to the radial star formation profile since \nuc{26}{Al} 
nucleosynthesis may depend on metallicity.
It will be important to determine this metallicity dependence in order 
to extract the true star formation profile from gamma-ray line data.
Valuable information about the metallicity dependence will come from a 
precise comparison of the 1.809 \MeV\ longitude profile to the profile of 
free-free emission \cite{knoedl99a}.
Additionally, observations of gamma-ray lines from \nuc{60}{Fe}, an 
isotope that is only believed to be produced during supernova 
explosions, can help to distinguish between hydrostatically and 
explosively produced \nuc{26}{Al}, and therefore help to disentangle 
the metallicity dependencies for the different candidate sources.

%%% Table: COMPTEL 26Al data %%%%%%%%%%%%%%%%%%%%%%%%%%%%%%%%%%%%%%%%%%%%%%%%%
%%%%%%%%%%%%%%%%%%%%%%%%%%%%%%%%%%%%%%%%%%%%%%%%%%%%%%%%%%%%%%%%%%%%%%%%%%%%%%

%%%%%%%%%%%%%%%%%%%%%%%%%%%%%%%%%%%%%%%%%%%%%%%%%%%%%%%%%%%%%%%%%%%%%%%%%%%%%%
% Annihilation line
%%%%%%%%%%%%%%%%%%%%%%%%%%%%%%%%%%%%%%%%%%%%%%%%%%%%%%%%%%%%%%%%%%%%%%%%%%%%%%
\section{Annihilation line}

The 511 keV gamma-ray line due to annihilation of positrons and 
electrons in the interstellar medium has been observed by numerous 
instruments (see \cite{harris98} and references therein).
At least two galactic emission components have been identified so far:
an extended bulge component and a disk component.
Indications of a third component situated above the galactic centre 
have been reported \cite{purcell97,harris98}, yet still needs
confirmation by more sensitive instruments.

The galactic disk component may be explained by radioactive positron 
emitters, such as \nuc{26}{Al}, \nuc{44}{Sc}, \nuc{56}{Co} 
\cite{lingenfelter89}.
Although all these isotopes are also gamma-ray line emitters,
only \nuc{26}{Al} will lead to correlated gamma-ray line and 511 keV 
emission since the typical annihilation time scale of some $10^{5}$ 
yrs considerably exceeds the lifetime of the other isotopes.
Consequently, 511 keV line-emission is a potential tracer of extinct 
short-lived galactic radioactivities.

The origin of the galactic bulge component is much less clear.
Reports of time-variable, possibly red-shifted 511 keV line features from 
the galactic centre direction led to the idea of compact objects being 
responsible for the galactic bulge component \cite{lingenfelter89}.
However, recent observations using more sensitive instruments could 
not confirm any time-variability \cite{jung95,smith96,harris98,cheng98}.
Actually, the most plausible source of the galactic bulge component 
may also be extinct short-lived radioisotopes from an old stellar 
population with a prominent galactic bulge component.
Type Ia supernovae could be good candidates for such a population since 
they produce appreciable amounts of \nuc{56}{Co} (a positron emitter)
and they are believed to belong to the old stellar population.
Hence it may turn out that the 511 keV annihilation line is an 
excellent tool fur studies of the galactic SN Ia population. 

Some information about the annihilation environment is obtained from 
511 keV line shape measurements and the determination of the positronium 
fraction.
In fact, positrons and electrons may eventually form a short-lived
hydrogen-like system called positronium which decays either into
two 511 keV photons or a three photon continuum below 511 keV.
The relative intensities of both components carry information about 
the fraction $f$ of annihilations via positronium formation,
probing the thermodynamic and ionisation state of the annihilation
environment \cite{guessom91}.
Recent observations suggest a positronium fraction of $f=0.9-1.0$ 
for the galactic bulge component \cite{kinzer96,harris98}.
Together with the only moderately broadened 511 keV line width, this 
indicates that annihilation mainly occurs in the warm 
neutral or ionised interstellar medium \cite{harris98}.

%%%%%%%%%%%%%%%%%%%%%%%%%%%%%%%%%%%%%%%%%%%%%%%%%%%%%%%%%%%%%%%%%%%%%%%%%%%%%%
% Perspectives
%%%%%%%%%%%%%%%%%%%%%%%%%%%%%%%%%%%%%%%%%%%%%%%%%%%%%%%%%%%%%%%%%%%%%%%%%%%%%%
\section{Perspectives}

Due to continuing progress in instrumentation, the field of gamma-ray 
line astronomy has now become a new complementary window to the 
universe.
With the COMPTEL and OSSE telescopes on {\em CGRO}, the entire sky 
has been imaged for the first time in the light of gamma-ray lines, 
leading to maps of 511 keV annihilation radiation and \nuc{26}{Al} 1.809 MeV 
emission.
New gamma-ray lines have been discovered, such as the 1.157 MeV line 
from \nuc{44}{Ti} or several decay lines from \nuc{56}{Co} and 
\nuc{57}{Co}.
Gamma-ray lines probe aspects of nucleosynthesis, stellar evolution, and 
supernova physics that are difficult to access by other means.
Additionally, they provide tracers of galactic activity and improve 
our understanding of the interstellar recycling processes.

The progress will continue.
In 2001, ESA's {\em INTEGRAL} gamma-ray observatory will be launched which 
is equipped with two gamma-ray telescopes, optimised for high-resolution 
imaging (IBIS) and high-resolution spectroscopy (SPI) 
(see {\tt http://astro.estec.esa.nl/SA-general/Projects/Integral/ 
integral.html}).
Gamma-ray line astrophysics figures among the prime objectives of this 
mission.
With respect to precedent instruments, the {\em INTEGRAL} telescopes 
provide enhanced sensitivity together with improved angular and 
spectral resolution.
In particular, SPI will map gamma-ray lines with a spectral 
resolution $E / \Delta E \sim 500$, corresponding to Doppler velocities 
of $\sim 600$ km s$^{-1}$.
It will provide much more detailed maps of the galactic 511 keV and 
1.809 MeV line emissions and determine their line profiles with 
unprecedented accuracy.
Following nucleosynthesis theory, the detection of diffuse galactic
gamma-ray line emission from \nuc{60}{Fe} decay is expected, and more
\nuc{44}{Ti} supernova remnants should be discovered.

Further progress is expected from new instrumental concepts, such as 
a high-resolution Compton telescope (eg.~\cite{johnson95,aprile98})
or a crystal lens diffraction telescope \cite{ballmoos96}.
The latter concept has the outstanding capacity of providing extremely 
high signal-to-noise ratios, leading to unprecedented sensitivities 
to gamma-ray lines.
A first balloon flight for a gamma-ray lens prototype is scheduled for 
summer 2000
(see {\tt http://www.cesr.fr/$\sim$pvb/Claire/index.htlm}).
Potentially, a space borne gamma-ray lens telescope may observe an 
extragalactic Type Ia supernova every few months, making gamma-ray 
line observations a standard tool for supernova research.
Such observations will help to improve our understanding of the 
supernova phenomena, which is particularly important if supernovae 
shall be used as standard candles to probe cosmology.

%%%%%%%%%%%%%%%%%%%%%%%%%%%%%%%%%%%%%%%%%%%%%%%%%%%%%%%%%%%%%%%%%%%%%%%%%%%%%%%%
% References
%%%%%%%%%%%%%%%%%%%%%%%%%%%%%%%%%%%%%%%%%%%%%%%%%%%%%%%%%%%%%%%%%%%%%%%%%%%%%%%%

\end{document}